\documentstyle[prl,aps, epsfig, multicol]{revtex}
\begin{document}
\draft
\title{Onset of sliding friction in 
incommensurate systems}
\author{L. Consoli, H. J. F. Knops, and A. Fasolino}
\address{Institute for Theoretical Physics, University of Nijmegen, 
Toernooiveld 1, 6525 ED Nijmegen, The Netherlands}
\date{\today}
\maketitle
\begin{abstract}
We study the dynamics of an  incommensurate chain sliding on a 
periodic lattice, 
modeled by the Frenkel Kontorova hamiltonian with initial kinetic energy,
without damping and driving terms. 
We show that the onset of friction is due to a novel kind of
dissipative parametric resonances, 
involving several resonant phonons which are driven by 
the (dissipationless) coupling of the center of mass 
motion to the phonons  with wavevector related to the 
modulating potential. We establish quantitative estimates for 
their existence in finite systems and
point out the analogy with the induction phenomenon in Fermi-Ulam-Pasta 
lattices. 

\end{abstract}

\begin{multicols}{2}
The possibility of measuring friction at the atomic level provided by 
the Lateral Force Microscopes\cite{Mate} and Quartz Crystal
Microbalance\cite{Krim} has stimulated intense research on 
this topic\cite{book}. Phonon excitations are the dominant cause 
of friction in many cases\cite{Tomm}.
Most studies are carried out for one-dimensional 
non-linear lattices\cite{shinjo,Aubry2,Braun,StrElm,Bambi,Sokoloff,Braim,Hent} 
and in
particular for the Frenkel-Kontorova (FK) model\cite{FK},
where the surface layer is modeled by a harmonic chain and the substrate is 
replaced by a rigid periodic modulation potential. 
The majority\cite{Aubry2,Braun,StrElm,Bambi,Sokoloff,Braim,Hent} examines the 
steady state of the dynamical FK model in presence of dissipation representing
the coupling of phonons to other, undescribed degrees of freedom.\linebreak
\hspace*{0.3cm}We study the dynamics of an undriven incommensurate FK chain.
Our aim is to ascertain whether the experimentally observed 
superlubricity \cite{Hirexp} can be due to the blocking of the phonon
channels caused by an incommensurate contact of the two sliding surfaces.
Therefore we do not include any explicit damping of the phonon modes, 
since we wish to find out if they can be excited at all by the
motion of the center of mass (CM). In an earlier study, Shinjo and Hirano\cite{shinjo} 
found a superlubric regime for this model, where the chain would 
slide indefinitely without dynamic friction but with a recurrent exchange of
kinetic energy between CM and a single internal mode. We will show that their
finding is oversimplified by either too short simulation times or too small
system sizes. 
The inherent non-linear coupling of the CM to the phonons
leads to an irreversible decay of the CM velocity, albeit with very long time 
scales in some windows.
The dissipative mechanism is driven by the coupling of the CM to the modes 
with modulation wavevector $q$ or its harmonics, $\omega_{nq}$, 
and consists in a novel 
kind of parametric resonances with much wider
windows of instabilities than those deriving from the standard Mathieu 
equation\cite{handbook}. 
The importance of harmonic resonances at $\omega_{nq}$ has been 
pointed out before\cite{Aubry2,StrElm,Sokoloff}, with the 
suggestion\cite{Sokoloff} that  they 
could be absent in finite systems due to the 
discreteness of the phonon spectrum. However,  
it has not been realized that they act as a {\it driving} term for the
onset of dissipation via subsequent complex parametric 
excitations which we shall describe, establishing quantitative 
estimates for their existence in finite systems. 
A related mechanism has recently been identified in 
the resonant energy transfer in the induction phenomenon in Fermi-Ulam-Pasta 
lattices\cite{christie}. 

We start with the FK hamiltonian
\begin{equation}
{\cal H}=\sum_{n=1}^N\left[\frac{p_n^2}{2} + \frac{1}{2}\left(u_{n+1}-u_n-l
\right)^2 
+\frac{\lambda}{2\pi}\sin{ (\frac{2\pi u_n}{m} )}\right]
\label{Ham}
\end{equation}
where $u_n$ are the lattice positions and $l$ is the equilibrium spacing of 
the chain for $\lambda=0$,  
$\lambda$ being the strength of the coupling scaled to the 
elastic spring constant. We take an incommensurable ratio of 
$l$ to the period $m$ of the periodic potential, namely $m=1$, 
$l=(1+\sqrt 5)/2$. We consider chains of $N$ atoms with periodic boundary conditions
$u_{N+1}=Nl+u_1$.  Hence, in the numerical implementation,  
we have to choose commensurate approximations for $l$ so that
$l\times N=M\times  1$ with $N$ and $M$ integer, i.e. 
we express $l$ as ratio of consecutive Fibonacci numbers.
The ground state of this model displays the 
so-called Aubry transition\cite{Aubry}
from a modulated to a pinned configuration above a critical value 
$\lambda_c=0.14$.
Here we just note that in the limit of weak coupling 
($\lambda<<\lambda_c$),  
deviations from equidistant spacing $l$ in the ground state 
are modulated with the substrate modulation wavevector 
$q=2\pi l/m$\cite{modfun} as due to the frozen-in phonon $\omega_q$. 
Higher harmonics $nq$ have amplitudes which scale with $\lambda^n$.

We define the CM position and velocity as $Q= \frac {1}{N}\sum_nu_n$, 
$P= \frac {1}{N}\sum_n p_n$. By writing
$u_n=nl+x_n+Q$, the equations of motion for the deviations from a rigid 
displacement $x_n$ read
\begin{equation}
\ddot{x}_n=x_{n+1}+x_{n-1}-2x_n+\lambda\cos{(qn+2\pi x_n+2\pi Q)}
\label{eom1}
\end{equation}
We integrate by a Runge Kutta algorithm the $N$ Eqs. (\ref{eom1})
with initial momenta $p_n=P_0$ and  
$x_n(t=0)$ 
corresponding to the ground state.  For a given velocity $P$, 
particles pass over maxima of the potential with frequency 
$\Omega=2\pi P$, the so-called washboard frequency\cite{StrElm,Sokoloff}.
\begin{figure}
\epsfig{file=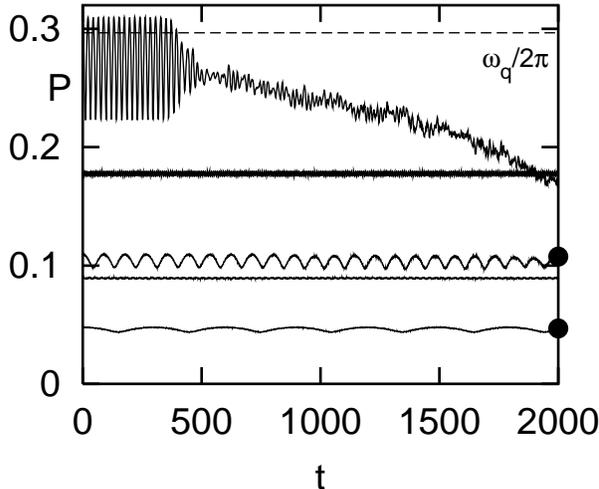, angle=-90}
\caption{Time dependence of the CM velocity for several values of $P_0$ for $N=144, 
\lambda=0.05\sim\frac{\lambda_c}{3}$. The dashed line corresponds to
            $P_0=\frac{\omega_q}{2\pi}=0.2966$. Close to higher resonances (solid dots) 
a similar oscillatory behavior is observed, accompanied by a slower decay which
is not apparent on the timescale of the figure.
}
\label{fig1}
\end{figure}
In Fig. 1 we show the time evolution of the CM momentum for 
$\lambda \sim \lambda_c/3$ and several values of $P_0$. 
According to  the phase diagram
of Ref.\onlinecite{shinjo} a superlubric behavior should be observed for 
this value of $\lambda$ and $P_0\geq 0.1$.
We find instead a non trivial time evolution with oscillations of varying 
period and amplitude and, remarkably, a very fast decay 
of the CM velocity for $P_0 \sim \omega_{q}/(2\pi)$ despite the absence of a 
damping term in Eq.~(\ref{eom1}). A similar, but much slower, decay
is found for $nP_0\sim \omega_{nq}/(2\pi)$. In the study of 
the driven underdamped FK\cite{StrElm} it is shown that, at
these superharmonic resonances, the differential mobility is extremely low.
Here, we work out an analytical description in terms of the phonon spectrum 
which explains this complex time evolution and  
identifies the dissipative mechanism which 
is triggered by these resonances.  
In the limit of weak coupling 
$\lambda$ it is convenient to go
from real to reciprocal space by defining Fourier transformed coordinates
$x_k=\frac{1}{N}\sum_n{e^{-ik n}x_n}$ and $x_n=\sum_k{e^{ik n}x_k}$, where 
$k=2\pi n/N$ and the normalization is chosen to remove the explicit
$N$-dependence in the equations of motion, which become:
\begin{mathletters}
\label{eom2}
\begin{eqnarray}
\ddot{x}_k&=&-\omega^2_k x_k + \frac{\lambda}{2N}\sum_n e^{-ikn}\left[e^{iqn} 
e^{i2\pi Q}e^{i2\pi x_n}  +c.c.\right]\label{xqgen}\\
\ddot{Q}&=&\frac{\lambda}{2N}\sum_n \left[e^{i2\pi Q}e^{iqn}e^{i2\pi x_n}  
+c.c.\right]
\end{eqnarray}
\end{mathletters}
with $\omega_k=2 \vert \sin(k/2)\vert$.
We expand Eq. (\ref{xqgen}) in $x_n$  as:
\end{multicols}
\begin{equation}
\ddot{x}_k=-\omega^2_k x_k + \frac{\lambda}{2}\sum_{m=0}^\infty 
\frac{(i2\pi)^m}{m!}\sum_{k_1\ldots k_m}\left[e^{i2\pi Q}x_{k_1}\cdots 
x_{k_m}\delta_{k_1+\cdots
+k_m,-q+k} + (-1)^m
e^{-i2\pi Q} x_{k_1} \cdots x_{k_m}\delta_{k_1+\cdots +k_m,q+k} \right]
\label{expans1}
\end{equation}

\begin{multicols}{2}
Since in the ground state the only modes present in 
order $\lambda$ are $x_q = x_{-q} = \lambda / 2\omega_q^2$
the CM is coupled only to these  modes up to 
second order in $\lambda$:
\begin{mathletters}
\label{Qxq}
\begin{eqnarray}
\ddot{Q}&=&i\lambda\pi \left(e^{i2\pi Q}x_{-q} - 
e^{-i2\pi Q}x_{q}\right)  \label{Q}\\
\ddot{x}_q &=&-\omega^2_q x_q + \frac{\lambda}{2} e^{i2\pi Q}  
\label{xq+}\\
\ddot{x}_{-q}&=& -\omega^2_q x_{-q}+ \frac{\lambda}{2} e^{-i2\pi Q} 
\label{xq-}
\end{eqnarray}
\end{mathletters}

In Fig.~2  we compare the behavior of 
$P(t)=\dot Q(t)$, obtained by solving the minimal set 
of Eqs. (\ref{Qxq}) with the appropriate initial 
conditions $Q(t=0)=0$, $P(t=0)=P_0$, $x_q(t=0)=\lambda/(2\omega_q^2)$, 
$\dot x_q(t=0)=0$, with 
the one  obtained from the full system of Eqs.~(\ref{eom1}).
Eqs.~(\ref{Qxq}) reproduce very well the initial behavior of the CM velocity 
which displays oscillations of frequency $\Delta$ around the value 
$\Omega/2\pi$ but do not predict
the decay occurring at later times because, as we show next, this is 
due to coupling to other modes. To this aim, we analyze  the relation 
between the initial CM velocity $P_0$  and $\Omega/2\pi$, respectively 
$\Delta_{\pm}$.
\begin{figure}
\epsfig{file=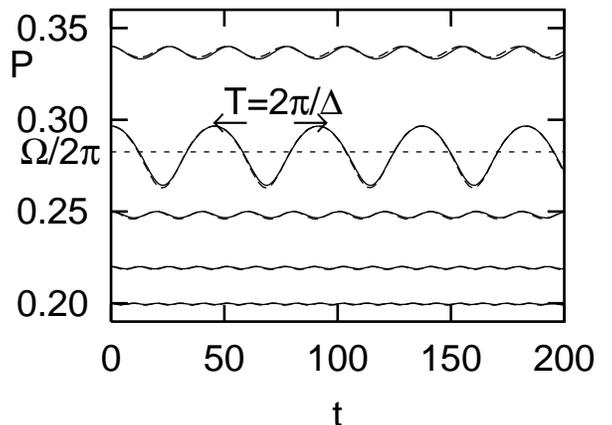, angle=-90}
\caption{Simulation of the full FK system according to 
Eq. (\ref{eom1}) (solid lines) and numerical solution of  Eqs. (\ref{Qxq}) (dashed lines) 
for $N=144$, $\lambda=0.015$, and several values of $P_0$ around $P_0=0.29$.
The differences between the two approaches are negligible. The average value of the CM velocity 
$\Omega/2\pi$ (horizontal dashed line) and the period of the oscillation for $P_0=0.29$ are also shown.
}
\label{fig2}
\end{figure}
Take as an ansatz for the CM motion:
\begin{equation}
Q(t)=\frac{\Omega}{2\pi}t+\alpha_+\sin(\Delta_+t)+\alpha_-\sin(\Delta_-t)
\label{ansatz}
\end{equation}
Inserting the ansatz (\ref{ansatz}) in the coupled set of Eqs. (\ref{Qxq}) keeping
only terms linear in $\alpha_{\pm}$, we find that both $\Delta_{\pm}$ are roots of:
\begin{equation} 
\Delta^2=\lambda^2 \pi^2 \left(2Z(0)-Z(\Delta)-Z(-\Delta)\right)
\label{eqDelta}
\end{equation}
$ Z(\Delta)$ being the impedance
\begin{equation} 
 Z(\Delta) = \frac{1}{\omega_{q}^2-(\Omega+\Delta)^2} \textrm{.}
\label{Delta}
\end{equation}
In general Eq. (\ref{eqDelta}) has (besides the trivial solution $\Delta=0$) 
indeed two solutions, related to the sum 
and difference of the two basic 
frequencies in the system, $\omega_q$ and $\Omega$ :
\begin{equation}
\Delta_\pm\cong|\omega_q\pm\Omega+\frac {\lambda^2 \pi^2} 
{2\omega_q(\Omega\pm\omega_q)^2} + \cdots|\label{deltam}
\end{equation}
Close to resonance, $\Omega\sim\omega_q$, the amplitude $\alpha_-$
dominates (see below) and the CM oscillates with a single frequency 
$\Delta=\Delta_-$ (see Fig.~2).
Very close to resonance (more precisely 
$\omega_q<\Omega<\omega_q+(2\lambda^2\pi^2/\omega_q)^{\frac{1}{3}}$),
the root $\Delta_-$ becomes imaginary, signaling an instability. 
In fact the system turns out to be bistable as it can be seen in Fig. 3 by the 
jump in $\Omega(P_0)$ as $P_0$ passes through $\omega_q/(2\pi)$.
\begin{figure}
\epsfig{file=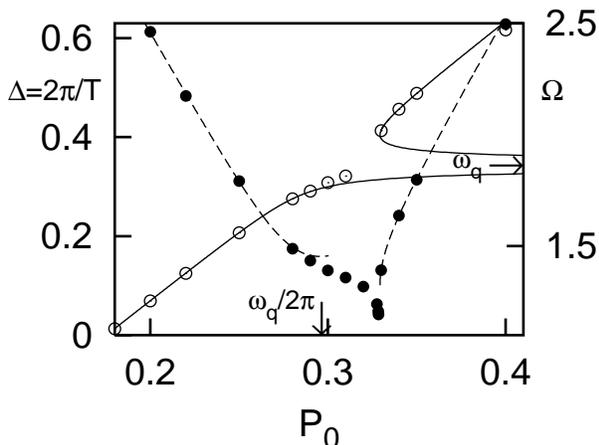, angle=-90}
\caption{Closed and open dots, frequencies $\Delta$ (left axis) and $\Omega$ (right axis) 
versus $P_0$ for simulations for $N=144$, $\lambda=0.015$. 
Dashed and solid lines: solutions of Eqs. (\ref{deltam}) and (\ref{p_0}) respectively.
}
\label{fig3}
\end{figure}
Analytically, the relation between $\Omega$ and $P_0$ and the
amplitudes $\alpha_{\pm}$, is determined
by matching the ansatz (\ref{ansatz}) with the initial condition:
\begin{equation}
\alpha_{\pm}= \frac{\lambda^2 \pi\Omega}{2\omega_q^2}\frac{1}{(\omega_q 
\pm\Omega)^3}
\label{alfa}
\end{equation}
\begin{equation}
P_0 = \frac{\Omega}{2\pi}+ \frac {\lambda^2\pi\Omega}{2\omega_q^2}\left[  
\frac{1}{(\Omega+\omega_q)^2}+\frac{1}{(\Omega-\omega_q)^2}\right] + \cdots
\label{p_0}
\end{equation}

The fact that Eq. (\ref{p_0}) has multiple solutions for $\Omega$ when $P_0 
\sim \omega_q/2\pi$ is in accordance with the jump seen in Fig. 3. However, 
Eq. (\ref{p_0}), which  is derived by keeping only linear terms in $\alpha_-$,
is not accurate enough to describe in detail the instability in the above 
range around $\omega_q$ where $\alpha_-$ diverges.

An initial behavior similar to that for $P_0\simeq\omega_q/2\pi$ is 
observed in Fig. 1 for $nP_0\simeq \omega_{nq}/n2\pi$. We examine
the case $n=2$.
Eq. (\ref{expans1}) shows
that $x_{2q}$ is driven in next order in $\lambda$ by $x_q$:
\begin{equation}
\ddot{x}_{2q}=-\omega_{2q}^2x_{2q} + i 2\lambda\pi e^{i2\pi Q}x_q
\label{x2q}
\end{equation}
When $2\pi Q\simeq \Omega t$, $x_q$ will be $\simeq \lambda e^{i\Omega t}$, 
so that
$x_{2q}$ is forced with amplitude $\lambda^2$ and frequency $2\Omega$, 
yielding 
resonance for $2\Omega=\omega_{2q}$. Since $x_{2q}$ couples back to 
$x_q$, we have
a set of equations similar to Eqs. (\ref{Qxq}), but at order $\lambda^2$.

We now come to the key issue, namely the onset of friction causing the decay 
of the CM velocity seen in Fig. 1 at later times, which
cannot be explained by the coupling of the CM to the main harmonics $nq$.  
Since $x_q$ is by far the largest
mode in the early stage, we consider second order terms involving $x_q$  
in Eq. (\ref{expans1}): 
\begin{equation}
\ddot{x}_k=- \left[ \omega^2_k +2 \lambda \pi^2 \left( e^{i2\pi Q} x_{-q} +
                                                     e^{-i2\pi Q} x_{q} \right)
                                       \right] x_k
\label{parres}
\end{equation}
Insertion of the solution obtained above for $x_q$ (Eq. (\ref{xq+})) and $Q$ 
(Eq. (\ref{ansatz})) yields
\begin{equation}
\ddot{x}_k=-\left[\omega^2_k + A + B\cos{(\Delta t)}\right]x_k
\label{ficata}
\end{equation}
with $A=2(\lambda\pi)^2/Z(0)$ and  $B\sim\alpha_-$.
Clearly, Eq. (\ref{ficata}) is a Mathieu parametric resonance for mode $x_k$.
The relevance of parametric resonances has been recently stressed\cite{Hent}. 
However here, due to the coupling of the CM 
to the modulation mode $q$, resonances are not with the washboard frequency
$\Omega$ but with $\Delta\sim\Omega-\omega_q$. Hence, we find 
instability windows around $\omega_k^2+A=(n\Delta/2)^2$.
Since $\Delta$ is small close to resonance, one expects to find instabilities 
for acoustic modes with $k$ small. Indeed, as shown in Fig. 4a, we find by 
solving Eq. (\ref{eom1}) that the decay of the CM is accompanied by the 
exponential increase of the  modes $k=2, 3, 4$ and, 
with a longer rise time, $k=1$. However, the instability windows 
resulting from Eq. (\ref{ficata}), shown in Fig. 4b, cannot 
explain the numerical results of Fig. 4a, i.e. the Mathieu formalism cannot 
explain the observed instability.
In Eq. (\ref{expans1}), the only linear terms left 
out in Eq. (\ref{parres})
are couplings with $x_{k\pm q}$, which are much higher order in $\lambda$. 
Nevertheless, these terms are crucial since they may cause new instabilities 
due to the fact that, for $k$ small, they  are 
also close to resonance.
We have solved the coupled set of equations for mode $x_{\pm k}$
and  $x_{k\pm q}$ :
\end{multicols}
\begin{mathletters}
\label{cfk}
\begin{eqnarray}
\ddot{x}_k & = & -\left[\omega^2_k +2\lambda\pi^2\left(e^{i2\pi Q}x_{-q}+
e^{-i2\pi Q}x_q\right)\right]x_k + i \lambda\pi\left(e^{i2\pi Q}
x_{k-q} - e^{-i2\pi Q}x_{k+q}\right) \label{cfk1} \\
\ddot{x}_{k\pm q} & = & -\left[\omega^2_{k \pm q} + 2\lambda\pi^2\left(
e^{i2\pi Q}x_{-q}+e^{-i2\pi Q}x_q\right)\right]x_{k\pm q} \pm i\lambda\pi
e^{\pm i2\pi Q}x_k \label{cfk2}
\end{eqnarray}
\end{mathletters}
\begin{multicols}{2}
together with Eqs. (\ref{Qxq}) for continuous $k$. Indeed, we find a wider 
range of instabilities, giving 
a detailed account of the
numerical result as shown in Fig. 4b. This mechanism where a parametric 
resonance is enhanced by coupling to near resonant modes  
is quite general in systems with a quasi continuous
spectrum of excitations and is 
related to the one proposed~\cite{christie} in explaining 
instabilities in the FPU  chain in a different physical context. \linebreak
\hspace*{0.3cm}The number of particles in the chain is an important parameter. When 
this number is very small, the chain is in fact commensurate and the phase
of the CM is locked (the gap scales as $\lambda^N$ due to Umklapp terms). Next, one
enters a stage of apparent superlubric behavior due to the fact that the
spectrum is still discrete on the scale of the size of the instability windows 
discussed above. For $N=144$ and $\lambda=\frac{1}{3}\lambda_c$ (Fig. \ref{fig1})
we only begin to see the decay for values of $P_0$ close to resonances.
The experimentally observed superlubricity in\cite{Hirexp} could then be due either
to the finiteness of the system or to the low sliding velocities.\linebreak
\hspace*{0.3cm}The above described multiple parametric excitation gives rise to an 
effective damping for the system via a cascade of couplings
to more and more modes via the non-linear terms in Eq. (\ref{expans1}).
It remains an open question if this mechanism will eventually lead to 
a full or partial equilibrium distribution of energy over the 
normal
modes\cite{Ruffo} although our preliminary results support the former 
hypothesis even at weak couplings.  \linebreak
\hspace*{0.3cm}In summary, we have described in detail the mechanism which gives rise to 
friction during the sliding of a harmonic system onto an incommensurate 
substrate. The onset of friction occurs in two steps: the resonant coupling 
of the CM to modes with wavevector related to the substrate 
modulation leads to long wavelength oscillations which in turn drive a complex
parametric resonance involving several resonant modes. This mechanism is 
robust in that it leads to wide instability windows and represents a quite 
general mechanism for the onset of energy transfer in systems with a quasi continuous
spectrum of excitations.  

We are grateful to Ted Janssen for many constructive discussions and 
for his support.
\begin{figure}
\epsfig{file=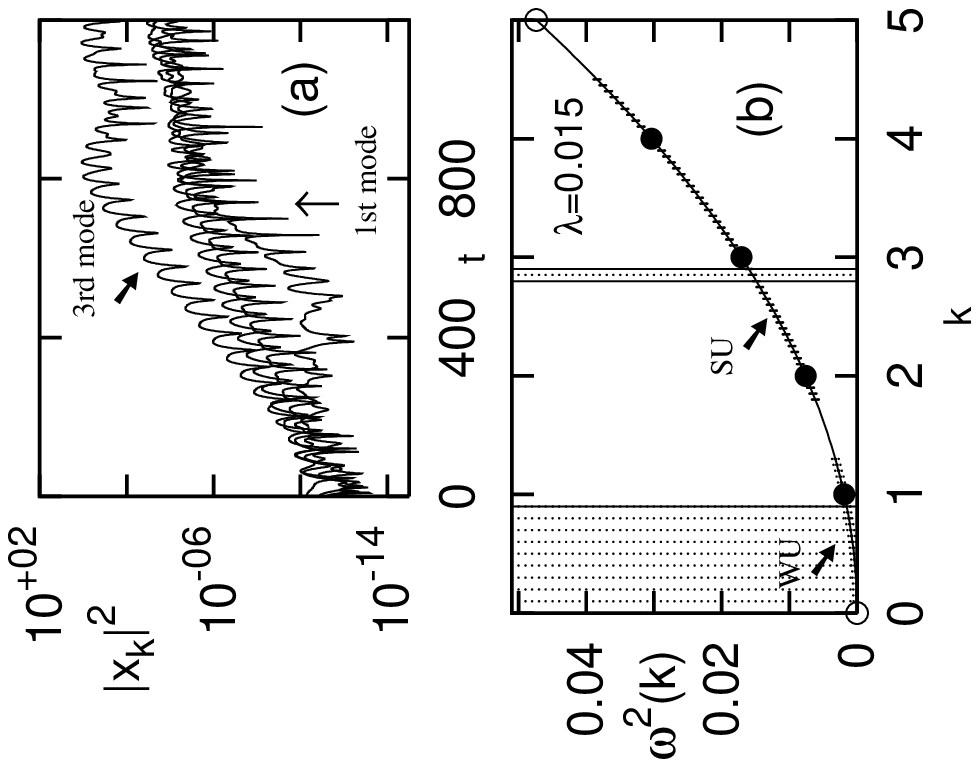, angle=-90}
\caption{(a) $|x_k(t)|^2$ of the first 4 modes from Eq. (\ref{eom1}) with 
$N=144$, $\lambda=0.015$ and $P_0=0.29$.
Note that the first mode has a longer rise time and that the third mode is the 
most unstable. (b) Dispersion relation for  a chain of $144$ atoms
($k$-values in units of $(\frac{2\pi}{144}$)). 
Unstable modes resulting from the full simulation are represented by solid 
dots. The shaded $k$-ranges give the instability windows resulting from the 
Mathieu-type Eq.(\ref{parres}) and cannot explain the simulation. 
Conversely the wiggled ranges of the phonon dispersion (WU=weakly unstable, 
SU=strongly unstable) are the instability windows predicted by Eqs. (\ref{Qxq}) and 
(\ref{cfk}).
They explain all instabilities as well as the long rise time of the 
first mode (WU) and the shortest one of the third mode which falls in the 
middle of the SU range.}
\label{fig4}
\end{figure}

\end{multicols}
\end{document}